\documentclass[12pt, amsfonts, amssymb,color]{article}

\usepackage{amsmath, amssymb, amsfonts, latexsym, bm, hyperref}
\def\R{\mathbb{R}}
\def\K{\mathbb{K}}
\def\C{\mathbb{C}}
\begin{document}

\renewcommand\theequation{\arabic{section}.\arabic{equation}}
\catcode`@=11 \@addtoreset{equation}{section}
\newtheorem{axiom}{Definition}[section]
\newtheorem{theo}{Theorem}[section]
\newtheorem{axiom2}{Example}[section]
\newtheorem{lem}{Lemma}[section]
\newtheorem{prop}{Proposition}[section]
\newtheorem{cor}{Corollary}[section]
\newcommand{\be}{\begin{equation}}
\newcommand{\ee}{\end{equation}}

\newcommand{\equal}{\!\!\!&=&\!\!\!}
\newcommand{\rd}{\partial}
\newcommand{\g}{\hat {\cal G}}
\newcommand{\bo}{\bigodot}
\newcommand{\res}{\mathop{\mbox{\rm res}}}
\newcommand{\diag}{\mathop{\mbox{\rm diag}}}
\newcommand{\Tr}{\mathop{\mbox{\rm Tr}}}
\newcommand{\const}{\mbox{\rm const.}\;}
\newcommand{\cA}{{\cal A}}
\newcommand{\bA}{{\bf A}}
\newcommand{\Abar}{{\bar{A}}}
\newcommand{\cAbar}{{\bar{\cA}}}
\newcommand{\bAbar}{{\bar{\bA}}}
\newcommand{\cB}{{\cal B}}
\newcommand{\bB}{{\bf B}}
\newcommand{\Bbar}{{\bar{B}}}
\newcommand{\cBbar}{{\bar{\cB}}}
\newcommand{\bBbar}{{\bar{\bB}}}
\newcommand{\bC}{{\bf C}}
\newcommand{\cbar}{{\bar{c}}}
\newcommand{\Cbar}{{\bar{C}}}
\newcommand{\Hbar}{{\bar{H}}}
\newcommand{\cL}{{\cal L}}
\newcommand{\bL}{{\bf L}}
\newcommand{\Lbar}{{\bar{L}}}
\newcommand{\cLbar}{{\bar{\cL}}}
\newcommand{\bLbar}{{\bar{\bL}}}
\newcommand{\cM}{{\cal M}}
\newcommand{\bM}{{\bf M}}
\newcommand{\Mbar}{{\bar{M}}}
\newcommand{\cMbar}{{\bar{\cM}}}
\newcommand{\bMbar}{{\bar{\bM}}}
\newcommand{\cP}{{\cal P}}
\newcommand{\cQ}{{\cal Q}}
\newcommand{\bU}{{\bf U}}
\newcommand{\bR}{{\bf R}}
\newcommand{\cW}{{\cal W}}
\newcommand{\bW}{{\bf W}}
\newcommand{\bZ}{{\bf Z}}
\newcommand{\Wbar}{{\bar{W}}}
\newcommand{\Xbar}{{\bar{X}}}
\newcommand{\cWbar}{{\bar{\cW}}}
\newcommand{\bWbar}{{\bar{\bW}}}
\newcommand{\abar}{{\bar{a}}}
\newcommand{\nbar}{{\bar{n}}}
\newcommand{\pbar}{{\bar{p}}}
\newcommand{\tbar}{{\bar{t}}}
\newcommand{\ubar}{{\bar{u}}}
\newcommand{\utilde}{\tilde{u}}
\newcommand{\vbar}{{\bar{v}}}
\newcommand{\wbar}{{\bar{w}}}
\newcommand{\phibar}{{\bar{\phi}}}
\newcommand{\Psibar}{{\bar{\Psi}}}
\newcommand{\bLambda}{{\bf \Lambda}}
\newcommand{\bDelta}{{\bf \Delta}}
\newcommand{\p}{\partial}
\newcommand{\om}{{\Omega \cal G}}
\newcommand{\ID}{{\mathbb{D}}}
\newtheorem{defi}{Definition}[section]
\newtheorem{exam}{Example}[section]
\def\R{\mathbb{R}}
\def\K{\mathbb{K}}
\def\C{\mathbb{C}}
\def\J{\mathbb{J}}
\def\X{\mathbb{X}}

\title{Jacobi-Maupertuis metric of Li\'{e}nard type equations \\and Jacobi Last Multiplier \\}
\author{ Sumanto Chanda$^1$, A.Ghose-Choudhury$^2$, Partha Guha$^{1,3}$ }

 \maketitle

\begin{minipage}{0.3\textwidth}
\begin{flushleft}
\textit{\small $ ^{1}$ S.N. Bose National Centre for Basic Sciences} \\
\textit{\small JD Block, Sector-3, Salt Lake, Kolkata-700098, INDIA.} \\
\texttt{\small sumanto12@bose.res.in \\  partha@bose.res.in}
\end{flushleft}
\end{minipage}
\begin{minipage}{0.3\textwidth}
\begin{center}
\textit{\small $^2$ Department of Physics, Surendranath  College} \\
\textit{\small 24/2 Mahatma Gandhi Road, \\ Kolkata - 700009, India.} \\
\texttt{\small aghosechoudhury@gmail.com}
\end{center}
\end{minipage}
\begin{minipage}{0.3\textwidth}
\begin{flushright}
\textit{\small $^3$ Instituto de F\'isica de S\~ao Carlos; IFSC/USP} \\
\textit{\small Universidade de S\~ao Paulo Caixa Postal 369,} \\
\textit{\small CEP 13560-970, \\ S\~ao Carlos-SP, Brazil}
\end{flushright}
\end{minipage}


\date{ }

 \maketitle

\begin{abstract}
We present a construction of the Jacobi-Maupertuis (JM) principle for an equation of the Li\'enard type, \textit{viz}
$\ddot{x} + f(x) \dot{x}^2 + g(x) = 0$ using Jacobi's last multiplier. The JM metric allows
us to reformulate the Newtonian equation of motion  for a variable mass as a geodesic equation for a Riemannian metric.
We illustrate the procedure with examples of Painlev\'e-Gambier XXI, the Jacobi equation and the Henon-Heiles
system.
\end{abstract}

\paragraph{Mathematics Classification (2000):} 34C14, 34C20.


\section{Introduction}
Nonlinear differential equations of the Li\'{e}nard type occupy a special place in the study of dynamical systems as they serve to model various physical, chemical and biological processes. The standard Li\'{e}nard equation involves a dissipative term depending linearly on the velocity. However there are practical problems in which higher order dependance on velocities are appropriate. Such equations have the generic form $\ddot{x}+f(x)\dot{x}^2+g(x)=0$. It is interesting to note that equations of this type naturally arise in Newtonian dynamics when the mass, instead of being a constant, is allowed to vary with the position coordinate-- the so called position dependent mass (PDM) scenario. There is also an alternative mechanism in which this dependance on a mass function manifests itself  in the context of differential systems, namely through Jacobi's last multiplier (JLM). The JLM originally arose in the problem of reducing a system of first-order ordinary differential equations to quadrature and has a long and chequered history. In recent years its role in the context of the inverse problem of dynamical systems has led to a revival of interest in the JLM. In this brief note we examine the connection between the JLM and the principle of least action within the framework of a Li\'{e}nard type differential equation with a quadratic dependance on the velocity.\\

\noindent
It is known that the Li\'enard type equation is connected to the Painlev\'e-Gambier equations \cite{CGK,GGCG}.
So it is natural
for us to ask whether we can reformulate the subclass of the Painlev\'e-Gambier family as geodesic equations
for a Riemannian metric using the Jacobi-Maupertuis principle.
There are several choices for a Riemannian manifold and metric tensor: a space-
time configuration manifold and the Eisenhart metric (for example, \cite{cgg,cgg1,Gibbons,SHS},
a configuration manifold and the Jacobi-Maupertuis metric \cite{nom,Pettini}. In this paper we choose a
configuration space of an analyzed system for a Riemannian manifold. The crux of the matter is that the Hamiltonian or energy function provided by the JLM should remain constant for these equations.\\

\noindent
\textbf{Main Result} Let ${\cal V}$ be a Hamiltonian vector field of the Li\'enard type equation $\ddot{x} + f(x)\dot{x}^{2} + g(x) = 0$ in ${\Bbb R}^2$ with Hamiltonian
$H = \frac{1}{2}M(x)\dot{x}^{2} + U(x)$, where $M(x) = exp(2\int^x f(s)ds)$ and
$U(x) = \int^x M(s)g(s)ds$. Then by Maupertuis principle,  ${\cal V}$ coincides with the
trajectories of the modified vector field ${\cal V}^{\prime}$ on the fixed isoenergy level
$H(x,\dot{x}) = E$ for the Hamiltonian ${\tilde H} = \frac{1}{2(E - U(x))}M(x)\dot{x}^{2}$.
This defines a geodesic flow of some Riemannian metric given by Jacobi. In other words,
solutions to the Li\'enard type equation with energy $E$ are, after reparametrization, geodesics
for the Jacobi-Maupertuis metric.

\smallskip

\noindent
A {\it corollary} of the main result shows that we can reformulate the Newtonian equation
of motion for a variable mass, Painlev\'e-Gambier XXI equation, the Jacobi equation and Henon-Heiles system
in terms of geodesic flows of the Jacobi-Maupertuis metric.

\bigskip

\noindent
The outline of the letter is as follows: in section 2 we introduce the Jacobi Last Multiplier and point out its connection to the Lagrangian of a second-order ODE. Thereafter we explicitly derive the Lagrangian and the Hamiltonian functions for a Li\'{e}nard equation of the second kind, i.e., with a quadratic dependance on the velocity and highlight the role of the position dependant mass term. In section 3 we express the equation in terms of geodesic flows of the Jacobi-Maupertuis metric and some observations regarding the geometric  consequences of the PDM are outlined. Explicit examples from the Painlev\'{e}-Gambier family of equations are considered along with the two-dimensional Henon-Heiles system.\\
\section{Lagrangians and the Jacobi Last Multiplier }
 Let $M=M(x^1,...,x^n)$ be a non-negative $C^1$ function
 non-identically vanishing on any open subset of $\R^n$, then $M$
 is a Jacobi multiplier of the vector field
 $\X=W^i\frac{\partial}{\partial
 x^i}$ if
\begin{equation}
\int_DM(x^1,...,x^n)dx^1...dx^n=\int_{\phi_t(D)}
 M(x^1,...,x^n)dx^1...dx^n
\end{equation}
where $D$ is any open subset of $\R^n$ and $\phi_t(.)$ is the flow
generated by the solution ${\bm x}= {\bm x}(t)$ of the system
\begin{equation}
\frac{dx^i}{dt} = W^i(x^1,...,x^n) \qquad i=1,...,n.
\end{equation}
 Thus the Jacobi multiplier can be viewed as the density
 associated with the invariant measure $\int_D M dx$.
The divergence free condition is
\begin{equation}
\frac{dM}{dt}+\frac{\partial W^i}{\partial x^i} M=0.
\end{equation}
The appellation `last' is a  historical legacy.
 If a Jacobi multiplier is known together with $(n-2)$ first
 integrals, we can reduce locally the $n$ dimensional system to a
 two-dimensional vector field on the intersection of the $(n-2)$
 level sets formed by the first integrals. The existence of a Jacobi
 Last Multiplier \cite{CG} then implies  the existence of an extra first
 integral and the system may therefore be reduced to quadrature. \\

 \noindent
 For a second-order ODE:
\begin{equation}\label{eom}
\ddot{x} = F(x,\dot{x},t) \qquad \Rightarrow \qquad \dot{x} = y, \ \dot{y} = F(x,y,t).
\end{equation}
we have
\begin{equation}
\label{dfree1} \frac{dM}{dt} + \frac{\partial F}{\partial y} M = 0.
\end{equation}
On the other hand by expanding the Euler-Lagrange equation of motion
\begin{equation}\label{EL1}
\frac{\partial L}{\partial x} - \frac{d}{dt} \left(\frac{\partial L}{\partial \dot x}\right) = 0,
\end{equation}
we have
$$\frac{\partial L}{\partial x} = {\dot y} \left(\frac{\partial^2 L}{\partial \dot x^2}\right) + {\dot x} \frac{\partial }{\partial x} \left(\frac{\partial L}{\partial \dot x}\right) = {\dot y} \left(\frac{\partial^2 L}{\partial \dot x^2}\right) + y \frac{\partial }{\partial \dot x} \left(\frac{\partial L}{\partial x}\right).$$
 Differentiating it w.r.t., $\dot{x}=y$, gives
\[ \begin{split}
\frac{\partial \ }{\partial \dot x} \left( \frac{\partial L}{\partial x} \right) &= \frac{\partial \dot y}{\partial y} \left(\frac{\partial^2 L}{\partial \dot x^2}\right) + \dot y \left(\frac{\partial^3 L}{\partial \dot x^3}\right) + \frac{\partial }{\partial \dot x} \left(\frac{\partial L}{\partial x}\right) + y \frac{\partial^2 \ }{\partial \dot x^2} \left(\frac{\partial L}{\partial x}\right), \\
\Rightarrow \qquad &\frac{\partial F}{\partial y} \left(\frac{\partial^2 L}{\partial \dot x^2}\right) + \left[ \dot y \frac{\partial \ }{\partial \dot x} \left( \frac{\partial^2 L}{\partial \dot x^2} \right) + y \frac{\partial \ }{\partial x} \left(  \frac{\partial^2 L}{\partial \dot x^2} \right) \right] = 0.
\end{split} \]
\begin{equation}
\label{dfree2} \therefore \qquad \frac{d}{dt}\left(\frac{\partial ^2 L}{\partial \dot{x}^2}\right) + \left(\frac{\partial F}{\partial y}\right) \left(\frac{\partial ^2 L}{\partial \dot{x}^2}\right) = 0.
\end{equation}
Thus, by comparing (\ref{dfree2}) to (\ref{dfree1}), we may identify the JLM as the following:
\begin{equation}
\label{density} M=\frac{\partial ^2 L}{\partial \dot{x}^2}.
\end{equation}
Given a JLM we can easily integrate (\ref{density}) twice to obtain
\begin{equation}
\label{lagsol} L(x,\dot{x},t) = \int^{\dot{x}} \left(\int^y M dz\right) dy +R(x,t) \dot{x}+S(x,t).
\end{equation}
where $R$ and $S$ are functions arising from integration.
 To determine these functions we substitute the
 Lagrangian  of (\ref{lagsol}) into the Euler-Lagrange equation of motion (\ref{EL1})  and
 compare the resulting equation with the given ODE  (\ref{eom}). \\ \\
Consider  now a Li\'enard equation of the second kind, \textit{viz}
\be
\label{lien} \ddot{x}+f(x)\dot{x}^2 +g(x)=0,
\ee
where $f$ and $g$ are defined in a neighbourhood of $0 \in {\Bbb R}$. We assume that $g(0) = 0$, which says that
$O$ is a critical point, and $x g(x) > 0$ in a punctured neighbourhood of $0 \in {\Bbb R}$, which ensures that the origin
is a centre. \\ \\

\begin{prop}\label{P1}A Li\'{e}nard equation of the second kind, $\ddot{x}+f(x)\dot{x}^2 +g(x)=0$, admits a Hamiltonian of the form $H=1/2 M(x)\dot{x}^2 +U(x)$ which is a constant of motion where $M(x)$ is the Jacobi last multiplier and $U(x)$ is a potential function.
\end{prop}
\noindent
{\bf Proof}: From the definition  (2.5) of the last multiplier  it follows that for the equation under consideration
\begin{equation}
M(x)=\exp(2F(x)) \qquad \mbox{where} \qquad F(x)=\int^x f(s) ds.
\end{equation}
Consequently according to (\ref{lagsol}), we have
\begin{equation}
L = \frac12 M(x) \dot{x}^2 + R(x,t) \dot{x} + S(x,t).
\end{equation}
From the Euler-Lagrange equation one finds that the functions
   $R$ and $S$ must satisfy
   $$S_x - R_t = - M(x)g(x)$$
This gives us the freedom to set
   $S=G_t - U(x)$ and $R=G_x$ for some gauge function $G$, so that there exists a potential function $U(x)$ given by
\begin{equation}
U(x) = \int^x M(s) g(s) ds.
\end{equation}
The Lagrangian then has the following appearance
\begin{equation}
\label{lienlag} L=\frac12 M(x) \dot{x}^2 - U(x) + \frac{dG}{dt}.
\end{equation}
Clearly the total derivative term may be ignored and by means of the standard Legendre transformation the Hamiltonian
is given by
\begin{equation}
\label{ham} H = \frac12 M(x) \dot{x}^2 + U(x).
\end{equation}
It is now straight forward to verify that $dH/dt=0$ so that $H=E$(say) is a constant of motion. This complete the proof.$\bullet$\\ \\
From (\ref{ham}) it is evident  that the JLM, $M(x)$, plays the role of a variable
mass term. We can reduce the differential system to a unit mass problem by defining a
transformation $x\longrightarrow X=\int_0^x \sqrt{M(s)}ds=\psi(x)$
whence
\begin{equation}
\frac12 \dot{X}^2 + \int_0^{\psi^{-1}(X)}M(s)g(s) ds = E.
\end{equation}
In terms of $X$ the equation of motion is given by
\begin{equation}
\ddot{X} + e^{F(\psi^{-1}(X))} g(\psi^{-1}(X)) = 0.
\end{equation}
We  now proceed to cover some fundamentals regarding the Jacobi metric, and deduce it
for the Li\'enard equation. We mainly follow the Nair \textit{et al}. formalism of Jacobi-Maupertuis principle and
elaborate on it in the next section.

\section{Jacobi-Maupertuis metric and Li\'enard type equation}

When the Hamiltonian is not explicitly time dependent, i.e., $H=E_0$, a constant, then the solutions may be restricted to the energy surface $E=E_0$. Suppose $Q$ is a manifold with local coordinates $x=\{x^i\}, i=1,...,n$ and $x(\tau) \in Q\subseteq \R^n$ be a curve with $\tau\in[0, T]$. Let $T_xQ$ and $T_x^*Q$ be the tangent and cotangent spaces with velocity $\dot{x}(\tau)\in T_xQ\subseteq\R^n$ and momenta $p(\tau)\in T_x^*Q\subseteq\R^n$. Denote by $\gamma$ a curve in the manifold $Q$ parametrized by $t\in[a, b]$ with $\gamma(a)=x_0$ and $\gamma(b)=x_N$. The according to the Maupertuis principle among all the curves $x(t)$ connecting the two points $x_0$ and $x_n$ parametrized such that $H(x, p)=E_0$ the trajectory of the Hamiltons equation of motion is an extremal of the integral of action
\be\label{x0}\int_\gamma p dx=\int_\gamma p \dot{x} dt=\int_\gamma \frac{\partial L(t)}{\partial\dot{x}} \dot{x}(t) dt.\ee
Here $L$ is assumed to be a regular Lagrangian $L:TQ\rightarrow \R$ where $L=K-U$ and the kinetic energy $K:TQ\rightarrow \R$.\\ \\

\begin{prop}Let the Hamiltonian $H=K+U$ be a constant of motion i.e., $H=E$(say) with the kinetic energy $K$ being a homogeneous quadratic function of the  velocities and $U(x)$ is some potential function such that $U(x)<E$: then there exists a Riemannian metric defined by $d\widetilde{s}=\sqrt{2(E-U(x))}ds$ with $K=1/2(ds/dt)^2$ such that the trajectories are the geodesic equations corresponding to the Jacobi-Maupertuis principle of least action.
\end{prop}
\noindent
{\bf Proof:} Let $ds^2$ be a Riemannian metric on the configuration space with kinetic energy
\be\label{x1} K=\frac 12 g_{ij}(x) \dot{x}^i \dot{x}^j=\frac 12\left(\frac{ds}{dt}\right)^2.\ee
As the total energy is a constant $E$ with potential $U(x)<E$ the Hamiltonian satisfies $H=K+U=E$. Because $K$ is a homogeneous quadratic function hence Euler theorem implies $ 2K = \dot{x}^i {\partial L}/{\partial \dot{x}^i}=(ds/dt)^2$. Therefore from (\ref{x0}) we have
$$\int_\gamma \frac{\partial L(t)}{\partial\dot{x}} \dot{x}(t) dt=\int_\gamma 2K dt=\int_\gamma 2K\frac{ds}{\sqrt{2K}}=\int_\gamma \sqrt{2K} ds=\int_\gamma \sqrt{2(E-U(x))}ds=\int_\gamma d\widetilde s,$$
where the Riemannian metric $\widetilde s$ is defined by $d\widetilde s=\sqrt{2(E-U(x))}ds$.
This shows that it is possible to derive a metric which is given by the kinetic energy itself \cite{cgg} and the trajectories are geodesics in the metric $d\widetilde s$. From (\ref{x1}) one finds $ds=\sqrt{q_{ij} dx^idx^j}$ and the Maupertuis principle involves solving for the stationary points of the action $\int \sqrt{2K} ds$, i.e.,
\be\label{x3}\delta\int \sqrt{2K} ds=0\;\;\;\mbox{or}\;\;\delta\int \sqrt{2(E-U(x))g_{ij} dx^idx^j}=0,\ee
with the integral being over the generalized coordinates $\{x^i\}$ along all paths connecting $\gamma(a)$ and $\gamma(b)$. \\

\noindent
It is evident from $d\widetilde s=\sqrt{2(E-U(x))g_{ij} dx^idx^j}$ that
\be\label{x4}d{\widetilde s}^2=\widetilde{g}_{ij} dx^idx^j\;\;\;\mbox{ where}\;\; \widetilde{g}_{ij}(x)=2(E-U(x))g_{ij}(x).\ee

\noindent
The geodesic equation corresponding to the least action $\delta \int_{s_1}^{s_2} dt \sqrt{\widetilde{g}_{ij} \dot x^i \dot x^j} = 0$
is given by
\be\label{x5}\frac{d^2x^i}{d \widetilde s^2} + \Gamma^i_{jk}\frac{dx^j}{d \widetilde s}\frac{dx^j}{d \widetilde s} = 0, \qquad \hbox{ where } \,\,\,\,\,
 \Gamma^i_{jk} = \frac12 \widetilde{g}^{il}\bigg(\frac{\partial \widetilde{g}_{jl}}{\partial x^k} + \frac{\partial \widetilde{g}_{kl}}{\partial x^j}
- \frac{\partial \widetilde{g}_{jk}}{\partial x^l}\bigg).\ee
This complete the proof. $\bullet$\\ \\
 For an equation of the Li\'{e}nard type given by (\ref{lien}) we have from Proposition (\ref{P1})
$$K=\frac{1}{2}M(x) \dot{x}^2\;\;\;\mbox{where}\;\;\;M(x)=\exp(2F(x))$$ so that $g_{11}(x)=M(x)$ while from the
Jacobi-Maupertuis (JM) metric  (\ref{x4}) we observe that $\widetilde{g}_{11}=2(E-U(x))M(x)$. The geodesic equation (\ref{x5}) therefore reduces to
$$\frac{d^2x}{d \widetilde s^2} + \Gamma^1_{11}\left(\frac{dx}{d\widetilde s}\right)^2=0\;\;\;\mbox{with}\;\;\;\Gamma^1_{11}=\frac{M^\prime(x)}{2M(x)}-\frac{U^\prime(x)}{2(E-U(x))},$$
or in explicit terms
\be\label{x6}\frac{d^2x}{d \widetilde s^2}+\left(\frac{M^\prime(x)}{2M(x)}-\frac{U^\prime(x)}{2(E-U(x))}\right)
\left(\frac{dx}{d\widetilde s}\right)^2=0.\ee
Eqn. (\ref{x6}) gives the geodesic for the JM-metric of a Li\'{e}nard equation of the type (\ref{lien}).\\ \\

\begin{prop}
 The geodesic equation (\ref{x6}) and (\ref{lien}) are equivalent.\end{prop}
\noindent
{\bf Proof:} From $K=E-U(x)=1/2 M(x)\dot{x}^2$  we have
\be\label{x6a}\dot{x}^2=2(E-U(x))/M(x)\;\;\; \mbox{ and as}\;\;\;
d\widetilde{s}^2=\widetilde{g}_{11} dx^2=2((E-U(x))M(x) dx^2,\ee it follows that
\be\label{x7}\frac{d\widetilde{s}}{dt}=2(E-U(x))\;\;\Rightarrow \frac{dx}{dt}=2(E-U(x))\frac{dx}{d\widetilde{s}}.\ee
This enables us to obtain
\begin{align}\frac{d^2 x}{d\widetilde{s}^2} &=\frac{1}{2(E-U(x))}\frac{d}{dt}\left\{\frac{1}{2(E-U(x))}\frac{dx}{dt}\right\}\nonumber\\
&=\frac{1}{4(E-U(x))^2}\left[\frac{d^2 x}{dt^2}+\frac{U^\prime(x)}{(E-U(x))}\dot{x}^2\right]\end{align}
Consequently (\ref{x6}), taking (\ref{x6a}) into account, assumes the form
$$\frac{1}{4(E-U(x))^2}\left[\frac{d^2 x}{dt^2}+\frac{U^\prime(x)}{(E-U(x))}\dot{x}^2\right]=
\left[\frac{U^\prime(x)}{2(E-U(x))}-\frac{M^\prime(x)}{2M(x)}\right]\frac{1}{4(E-U(x))^2}\dot{x}^2,$$
that is in other words we have
 \be\label{x8}\frac{d^2 x}{dt^2}+\frac{M^\prime(x)}{2M(x)}\dot{x}^2 +\frac{U^\prime(x)}{2(E-U(x))}\dot{x}^2=0.\ee
However as $\dot{x}^2=2(E-U(x))/M(x)$ the last term of the above equation can be expressed as $U^\prime(x)/M(x)$ and as a result the equation has the appearance
\be\label{x9}\frac{d^2 x}{dt^2}+\frac{M^\prime(x)}{2M(x)}\dot{x}^2 +\frac{U^\prime(x)}{M(x)}=0.\ee
 This equation reduces to (\ref{lien}) upon making the identifications $M(x)=\exp(2F(x))$ which implies $M^\prime(x)/2M(x)=f(x)$ and $U(x)=\int^x M(y) g(y) dy$ which implies $U^\prime(x)/M(x)=g(x)$ where $g(x)$  refers to the forcing term of the Li\'{e}nard equation (\ref{lien}).$\bullet$\\ \\

\noindent
\textbf{Remark:} Finally it is interesting to note how (\ref{lien}) or equivalently (\ref{x9}) may be viewed geometrically. To this end we write (\ref{x9})  as
\be\label{x10}\frac{d^2 x}{dt^2}+\frac{M^\prime(x)}{2M(x)}\dot{x}^2 =-\frac{U^\prime(x)}{M(x)}\ee
and look upon the right hand side as an external force function. Restricting ourselves to the left hand side we consider a 1+1 dimensional line element of the form $ds^2=c^2dt^2-M(x)dx^2=c^2d\tau^2$ which yields the following geodesic equations for a free particle moving in this spacetime, namely
$$\frac{d^2x}{d\tau^2}+\frac{M^\prime(x)}{2M(x)}\left(\frac{dx}{d\tau}\right)^2=0, \;\;\;\frac{d^2t}{d\tau^2}=0.$$
These equations imply upon elimination of the proper time $\tau$ the left hand side of (\ref{x10})and the latter may be recast as
$$\frac{d}{dt}\left(M(x)\dot{x}\right)=\frac{M^\prime(x)}{2}\dot{x}^2.$$
Thus from a Newtonian perspective we see that the position dependent mass function $M(x)$ changes the geometry of spacetime in a manner such that the particle experiences an additional geometric force $f_G=M^\prime(x)\dot{x}^2/2$. However unlike the case when the PDM is also a function of time \cite{MH} the curvature of spacetime is flat because as a result of the  transformation $dX=\sqrt{M(x)}dx$ one has $ds^2=c^2dt^2-dX^2$  and the resulting geodesic equation of a free particle in this transformed spacetime is just $\frac{d^2X}{dt^2}=0$ or
$$\frac{d}{dt}\left(\sqrt{M(x)}\frac{dx}{dt}\right)=0, \;\;\;\mbox{or}\;\;\;\frac{1}{2}M(x)\dot{x}^2=const.$$
which implies the conservation of the kinetic energy.\\

\noindent
We end this letter with a few examples for the purpose of illustration.

\noindent
\textbf{Example 1:} Painl\'{e}ve-Gambier XXI\\
 $$\ddot{x}-\frac{3}{4x}\dot{x}^2-3x^2=0$$
For this equation we have $F(x)=-3/4\int dx/x=-3/4\log|x|$ so that $M(x)=|x|^{-3/2}$ and as $2K=M(x)\dot{x}^2=g_{11}(x)\dot{x}^2$ we have $g_{11}(x)=M(x)=|x|^{-3/2}$ while $U(x)=\int^x M(z)g(z) dz=\mp 2x^{3/2}$ depending on whether $x>0$ or $x<0$. As a result we find have $\widetilde{g}_{11}=2(E\pm 2x^{3/2})|x|^{-3/2}$.\\

\noindent
\textbf{Example 2:} Jacobi equation\\
 $$\ddot{x}+\frac{1}{2}\phi_x \dot{x}^2+\phi_t\dot{x}+B(t, x)=0$$
Here $M(x,t)=\exp(\phi(x,t))=g_{11}$ and the Lagrangian is given by
$$L=\frac{1}{2}e^\phi \dot{x}^2-U(x,t), \;\;\mbox{where}\;\;U(x,t)=\int^x e^{\phi(y,t)}B(y,t) dy$$
It may be verified that the Hamiltonian is a constant of motion and $\widetilde{g}_{11}=2(E-U(x,t))\exp(\phi(x,t))$.
The geodesic equation is given by
$$\frac{d^2x}{d\widetilde{s}^2}+\Gamma^1_{11}\left(\frac{dx}{d\widetilde{s}}\right)^2=0, \;\;\mbox{with}\;\;
\Gamma^1_{11}=\frac{\phi_x}{2}-\frac{U_x}{2(E-U(x,t))}.$$

\noindent
\textbf{Example 3:} Henon-Heiles system\\
$$\ddot{x}=-(Ax+2\alpha xy)$$
$$\dot{y}=-(By+\alpha x^2-\beta y^2)$$
The above system has the Lagrangian
$$L(x,y, \dot{x}, \dot{y})=\frac{1}{2}(\dot{x}^2+\dot{y}^2)-\left(\frac{A}{2}x^2+\frac{B}{2}y^2+\alpha x^2 y-\frac{\beta}{3}y^3\right)$$
It is therefore easily seen that $M_{xx}=M_{yy}=1$ and it admits the first integral
$$I=\frac{1}{2}(\dot{x}^2+\dot{y}^2)+\left(\frac{A}{2}x^2+\frac{B}{2}y^2+\alpha x^2 y-\frac{\beta}{3}y^3\right),$$
which is just the Hamiltonian of the system. Consequently we have $g_{11}=M_{xx}=1$ and $g_{22}=M_{yy}=1$ while
$$\widetilde{g}_{11}=2(E-U(x,y))=\widetilde{g}_{22}, \;\;\mbox{where}\;\; U(x,y)=\frac{1}{2}(\dot{x}^2+\dot{y}^2)-\left(\frac{A}{2}x^2+\frac{B}{2}y^2+\alpha x^2 y-\frac{\beta}{3}y^3\right)$$
The geodesic equations have the following appearance:
$$\frac{d^2x}{d\widetilde{s}^2}-\frac{1}{2(E-U(x,y))}\left(U_x\left(\frac{dx}{d\widetilde{s}}\right)^2
+2U_y\left(\frac{dx}{d\widetilde{s}}\right)\left(\frac{dy}{d\widetilde{s}}\right)+ U_x\left(\frac{dy}{d\widetilde{s}}\right)^2\right)=0$$
$$\frac{d^2y}{d\widetilde{s}^2}-\frac{1}{2(E-U(x,y))}\left(U_y\left(\frac{dx}{d\widetilde{s}}\right)^2
+2U_x\left(\frac{dx}{d\widetilde{s}}\right)\left(\frac{dy}{d\widetilde{s}}\right)+ U_y\left(\frac{dy}{d\widetilde{s}}\right)^2\right)=0$$

\section*{Acknowledgement}
We are grateful to Professor Jaume Llibre for his valuable comments.
The research of PG is supported by FAPESP through
Instituto de Fisica de S\~ao Carlos, Universidade de Sao Paulo with grant number 2016/06560-6.

\end{document}